\RequirePackage{fix-cm}
\documentclass[smallextended]{svjour3}       
\smartqed  
\usepackage{bbding}
\usepackage{graphicx}
\usepackage{amssymb}
\usepackage{bm}
\newcommand{\bea}{\begin{eqnarray}}
\newcommand{\eea}{\end{eqnarray}}
\begin{document}

\title{Structure of the orbital excited $N^*$ from the Schwinger-Dyson equations
}


\author{K.~Raya  \and L.X. Guti\'errez \and A.~Bashir 
}


\institute{K. Raya (\Envelope) \at
             Departamento de F\'isica, Centro  de  Investigaci\'on y  de  Estudios  Avanzados
  del Instituto Polit\'ecnico Nacional.
Apartado Postal 14-740, 07000, Ciudad  de  M\'exico, M\'exico \\
              \email{khepani@ifm.umich.mx}      
           \and
           L.X. Guti\'errez \at
              CONACyT-Mesoamerican Centre for Theoretical Physics,
              Universidad Aut\'onoma de Chiapas, Carretera Zapata Km. 4, Real
              del Bosque (Ter\'an), Tuxtla Guti\'errez 29040, Chiapas,
              M\'exico \\
              \email{lxgutierrez@conacyt.mx}      
             \and
             A. Bashir \at
             Instituto de F\'isica y Matem\'aticas, Universidad Michoacana de San Nicol\'as de Hidalgo,
              Edificio C-3, Ciudad Universitaria, Morelia, Michoac\'an 58040, M\'exico \\
              \email{adnan@ifm.umich.mx}      
              }

\date{Received: date / Accepted: date}

\maketitle

\begin{abstract}

We present progress in the evaluation of $\gamma^* N \rightarrow N^*(1535)$ transition form factors 
in a quark-diquark picture of these baryons. Our analysis is based upon the fully-consistent 
treatment of a vector $\times$  vector contact interaction, embedded in the interlaced formalism of
Schwinger-Dyson and Bethe-Salpeter equations.

\end{abstract}

\section{Introduction}
\label{intro}

Dynamical chiral symmetry breaking (DCSB) and confinement are
emerging phenomena of quantum chromodynamics (QCD) not realized in
perturbation theory. Both phenomena have far reaching consequences 
in the spectrum and properties of hadrons. In fact, DCSB reflects itself directly in the mass
spectrum of hadrons. Chiral transformations rotate the composite
`scalar $\leftrightarrow$ pseudoscalar' as well as `vector
$\leftrightarrow$ axial vector' operators. If it were an exact
symmetry, we expect, for example, $\rho-a_1$ mass splitting to be
zero. However, the mass difference between low-lying parity
partners $\rho-a_1$, $N-N^*(1535)$ and $\Delta(1232)-\Delta(1600)$
all points towards DCSB of about 500-600 MeV. Therefore, the study
of transition form factors between parity partners such as $N
\rightarrow N^*(1535)$ provides us with a direct probe to
understand DCSB for light quarks. There has been considerable
experimental effort to study the helicity amplitudes for this
process, which can then be algebraically manipulated to yield the
transition form factors. The available data, for both the longitudinal ($S_{1/2}$) and transverse ($A_{1/2}$) helicity amplitudes, covers the region $Q^2\sim0-7$ GeV$^2$. MAID2007~\cite{Drechsel:2007if} reports helicity amplitudes for $Q^2$ up to $1.5$ GeV$^2$. Using the world data base of pion photo- and electroproduction and the data from Mainz, Bonn, Bates and JLab~\cite{Aznauryan:2009mx}, available at the time, further analysis by MAID~\cite{Tiator:2009mt} extracted longitudinal and transverse helicity amplitudes of nucleon resonance excitation $N^*(1535)$ to $Q^2 = 4.2$ GeV$^2$. In the JLab Hall C~\cite{Dalton:2008aa}, electroproduction of $\eta$ mesons in the $N^*(1535)$ resonance region was analyzed to extract $A_{1/2}$ for even higher photon virtuality, $Q^2 \sim 5.8, 7$ GeV$^2$, under the assumption that the $S_{1/2}$ contribution for the cross section is negligible\footnote{Although not detailed herein, our preliminary results indicate this to be a poor approximation.}. A value of $A_{1/2}$ at $Q^2 \sim 0$ is also reported by PDG~\cite{Beringer:1900zz}. The 12 GeV upgrade of the
JLab is likely extend the region of $Q^2$ to around 12 GeV$^2$. Thus time is ripe to make reliable predictions for large $Q^2$ against which
these forthcoming experimental results can be compared.

A symmetry preserving QCD treatment of Schwinger-Dyson equations (SDEs) is an ideal continuum framework to embark upon
this endeavor. Indeed, a study of static as well dynamic properties of all the ground state and excited mesons and baryons, with the minimum number of input parameters, has been a long term goal within this framework, see for example~\cite{Bashir:2012fs,Aznauryan:2012ba}. In this work, we outline our progress in the calculation of $\gamma^* N \rightarrow N^*(1535)$ transition form factors, using a symmetry-preserving treatment of a vector $\times$ vector contact interaction (CI).
The calculations described herein and those that are planned should provide useful benchmarks for the empirical results that will be extracted from modern data~\cite{Isupov:2017lnd}.


\section{The ingredients}
Quarks inside baryons tend to correlate into non-point-like diquarks~\cite{Lu:2017cln,Mezrag:2017znp,Chen:2017pse}. Therefore, it is reasonable to picture the baryon as a quark-diquark system,~\cite{Eichmann:2011vu}.
Each diquark has an associated meson. With minimal changes in the BSEs for mesons, one can infer the expressions of the corresponding equations for diquarks. One expects that ground-state positive-parity baryons are constituted almost exclusively by like parity diquarks, while their parity partners will likely involve diquarks of both positive and negative parity. Within this picture, several intermediate diquark transitions are required to evaluate the complete result. Calculation of $\gamma^* N \rightarrow N^*(1535)$ transition, {\em e.g.}, presupposes the
knowledge of the quark propagator, the Bethe-Salpeter amplitudes (BSA) of the mesons (and their corresponding diquarks) as well as the quark-photon electromagnetic interaction at different probing photon momenta.

\subsection{Quark Propagator}
The starting point is the quark propagator, which is obtained from the gap equation:
\begin{equation}
S^{-1}(p)=i\gamma \cdot p + m + \int \frac{d^4q}{(2\pi)^4}g^2 D_{\mu\nu}(p-q)\frac{\lambda^a}{2}\gamma_\mu S(q) \Gamma^a_\nu(q,p),
\label{eq:gap0}
\end{equation}
where $m$ is the current quark mass, $D_{\mu\nu}$ is the gluon propagator and $\Gamma_\nu^a$ is the fully-dressed quark-gluon vertex; both obey their own SDEs, {\em ad infinitum}. Therefore, we arrive at an infinite tower of coupled equations. The Green functions which directly enter the
gap equation are the gluon propagator and the quark-gluon vertex. There has been considerable progress towards understanding these Green functions, their
infrared enhancement, understanding the gluon mass scale and running coupling at large distances,~\cite{Cornwall:1981zr,Bogolubsky:2009dc,Ayala:2012pb,Bashir:2013zha,Bermudez:2017bpx,Binosi:2016nme}, connecting fundamental aspects
of QCD with hadron physics phenomenology,~\cite{Binosi:2014aea}.
A starting point which illustrates many of these key features of the strong interactions, is the contact interaction (CI)~\cite{GutierrezGuerrero:2010md,Roberts:2010rn,Wilson:2011aa,Bedolla:2015mpa,Raya:2017ggu}:
\begin{equation}
    g^2 D_{\mu\nu}(p-q)\to \delta_{\mu\nu}\frac{4\pi \alpha_{IR}}{m_G^2},\;\Gamma^a_\nu(q,p)\to \frac{\lambda^a}{2}\gamma_\nu,
    \label{eq:trunc}
\end{equation}
where $m_G = 0.8$ GeV is a gluon mass scale and $\alpha_{IR}=0.93\pi$ is commensurate with contemporary estimates of the zero-momentum running coupling of QCD. CI yields a momentum-independent mass function. Thus a general solution of Eq.~(\ref{eq:gap0}) is $S^{-1}(p)=i\gamma\cdot p + M$, where:
\begin{equation}
    M=m+M\frac{4 \alpha_{IR}}{3\pi m_G^2}\int_0^\infty ds\;\frac{s}{s+M^2}\;.
    \label{eq:mf}
\end{equation}
We perform proper time regularization,~\cite{Ebert:1996vx}:
\begin{eqnarray*}
\frac{1}{s+M^2}\to \int_{\tau_{IR}^2}^{\tau_{UV}^2} d\tau e^{-\tau(s+M^2)}= \frac{e^{-(s+M^2)\tau_{UV}^2}-e^{-(s+M^2)\tau_{IR}^2}}{s+M^2}\;,
\end{eqnarray*}
which guarantees confinement by ensuring the absence of quark production thresholds. $\Lambda_{IR}=1/\tau_{IR} = 0.24$ GeV and $\Lambda_{UV}=1/\tau_{UV}$ is an ultraviolet dynamical scale. It sets the scale of all dimensioned quantities because the theory is
non-renormalizable. Consequently, Eq.~(\ref{eq:mf}) becomes:
\begin{equation}
    M = m + M\frac{4 \alpha_{IR}}{3\pi m_G^2} \mathcal{C}^{iu}(M^2),
\end{equation}
where $\mathcal{C}^{iu}(M^2)/M^2= \Gamma(-1,M^2 \tau_{UV}^2)-\Gamma(-1,M^2 \tau_{IR}^2)$, with $\Gamma(\alpha,y)$ being the incomplete Gamma function.
\subsection{Bethe-Salpeter equation}
In the CI, with $\chi_H(q;P)=S(q+P)\Gamma_H(q;P)S(q)$, the quark-antiquark bound-state (meson) BSE is written as:
\begin{equation}
\label{eq:BSE}
    \Gamma_H(p;P) = -\frac{16 \pi}{3} \frac{\alpha_{IR}}{m_G^2}\int \frac{d^4q}{(2\pi)^4} \gamma_\mu \chi_H(q;P) \gamma_\mu\;. \label{eq:BSECI}
\end{equation}
The general tensor structure of $\Gamma_H(p;P)$ depends on the meson's spin (4 tensors for spin 0, 8 for spin 1). Due to the momentum-independent nature of the CI, we arrive at the following structures \cite{Lu:2017cln}:
\begin{eqnarray}
\Gamma_\pi(P) &=& \gamma_5 \left[ i E_\pi(P) + \frac{\gamma \cdot P}{M} F_\pi(P) \right]\;,\nonumber\\
\Gamma_\sigma(P) &=& \bm{1} \; E_\sigma(P)\;,\nonumber\\
\Gamma^{\rho}_{\mu}(P) &=& \gamma^T_{\mu} E_\rho(P)\;,\nonumber\\
\Gamma^{a_1}_{\mu}(P) &=& \gamma_5 \gamma^T_{\mu} E_{a_1}(P)\;,
\label{eq:BSE}
\end{eqnarray}
where $P_\mu \gamma_\mu^T = 0$. Crucially, $F_{\pi}$, the pseudovector component of the pion, cannot be neglected: it is an essential part of any consistent treatment of a momentum-independent quark-quark interaction.
For colored states (diquarks), $\Gamma_H^c(P) = \Gamma_H(P) C^\dagger H^c$, where $\vec{H}=\{i \lambda_c^7, -i \lambda_c^5, i \lambda_c^2 \}$ and $\lambda_c^k$ are the GellMann matrices; $C=\gamma_2\gamma_4$ is the charge conjugation matrix. Therefore, the right-hand-side of Eq.~(\ref{eq:BSE}) changes by a factor of $1/2$ when formulating a BSE for diquarks.
\\
\\
By setting $\Lambda_{UV}=0.905$ GeV and $m_{u/d} = m = 0.007$ GeV, one obtains a reasonably good description of $\pi$ and $\rho$ mesons: $m_{\pi,\rho}=0.140,0.929$ GeV, $f_{\pi,\rho}=0.101,0.129$ GeV, $M=0.368$ GeV \cite{Roberts:2010rn}.
\\
\\
In order to take into account spin-orbit repulsion for $L=1$ states, we include two additional phenomenological couplings \cite{Lu:2017cln}, $g_{SO}^{1,0}=0.25, 0.32$ for mesons and $g_{SO}^{1,0}\times 1.8$ for the corresponding diquarks (which are more loosely correlated), into the corresponding channels. One gets $m_{a_1}-m_{\rho} = 0.44$ GeV, in line with experiment. We summarize our results in the following table:
\begin{center}
\begin{table}[h!]
\centering
\caption{Computed meson (diquark) masses in GeV.}
\label{tab:1}       
\begin{tabular}{| c | c | c | c | c |}
\hline
& $\;\;\;\pi\;(0^+)\;\;\;$ & $\;\;\;\rho\;(1^+)\;\;\;$ & $\;\;\;\sigma\;(0^-)\;\;\;$ & $\;\;\;a_1\;(1^+)\;\;\;$ \\
\hline
$\;m_H\;$ & $0.14\;(0.78)$    & $0.93\;(1.06)$ & $1.22\;(1.15)$ & $1.37\;(1.33)$ \\
\hline
\end{tabular}
\end{table}
\end{center}
\subsection{Faddeev equation}
Our Faddeev kernel involves diquark breakup and reformation via exchange of a dressed-quark. We treat this exchange quark as in the static approximation~\cite{Buck:1992wz} $S(k) \to g_B^2/M$; $g_B = 1.18$ in order to produce nucleon (N) and $\Delta$ baryon masses inflated by 0.2 GeV \cite{Lu:2017cln}, since meson cloud effects are not included in this calculation.
\begin{figure}
\begin{center}
\vspace*{-3cm}
  \includegraphics[width=100mm,scale=0.5,angle=-90]{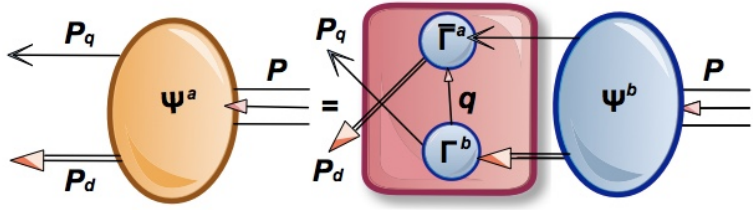}
\vspace*{-3cm}
\caption{Faddeev equation in quark-diquark picture. The rectangular shaded area demarcates the interaction kernel.}
\label{faddeev}       
\end{center}
\end{figure}


Diquarks inside $N$ and  $N^*(1535)$ correlate as follows: $0^\pm = [ud]^\pm$, $1^\pm = \{ud\}^\pm$. Additionally, we have $1^+ =\{uu\}^+$ for the charged and $1^+ =\{dd\}^+$ for the neutral states. Then, the Faddeev amplitudes are \cite{Lu:2017cln}:
\begin{eqnarray}
\nonumber
\psi^\pm u(P) &=& \Gamma_{0^+}^1 \Delta^{0^+}(K) \mathcal{S}^\pm(P)u(P) + \sum_{f=1,2} \Gamma_{1^+}^f \Delta_{\mu\nu}(K)\mathcal{A}_\nu^{\pm f} u(P)\\
\label{eq:fadAm}
&+& \Gamma_{0^-}\Delta^{0^-}(K) \mathcal{P}^\pm (P) u(P)+\Gamma_{1^-}\Delta^{1^-}_{\mu\nu}(K) \mathcal{V}_\nu^\pm (P) u(P),
\end{eqnarray}
where $u(P)$ are Dirac spinors and $\Delta^{0^{\pm},1^{\pm}}$ are standard diquark propagators \cite{Wilson:2011aa}. With the correlations:
\begin{eqnarray*}
\mathcal{S}^\pm(P)= \bm{1} \; s^\pm&,&\; i \mathcal{P}^\pm = p^\pm \gamma_5,\\
 \mathcal{A}^{\pm f}(P) = (a_1^{\pm f} \gamma_5 \gamma_\mu - i a_2^{\pm f} \gamma_5 \hat{P}_\mu)&,& i \mathcal{V}^\pm = (v_1^\pm \gamma_\mu - i v_2^\pm \bm{1} \hat{P}_\mu),
\end{eqnarray*}
and $\Psi_{(\mu)}u(P) = [\mathcal{S}^\pm(P),\;\mathcal{A}_\mu^{\pm f}(P),\;\mathcal{P}^\pm(P),\;\mathcal{V}_\mu^\pm(P)]u(P)$ ($\hat{P}^2=-1$), the Faddeev equation reads as:
\begin{equation}
    \Psi^T_{(\mu)} = -4 \int \frac{d^4q}{(2\pi)^4} \mathcal{M}_{(\mu,\nu)}(q,P) \Psi^T_{(\nu)}\;,
\end{equation}
where $\mathcal{M}_{(\mu,\nu)}(q,P)$ is the kernel, obtained from different projections of Eq.~(\ref{eq:fadAm}).
As with mesons and diquarks, spin-orbit repulsion is included by introducing a multiplicative factor, $g_{DB} = 0.1$, attached to diquarks whose parity is opposite to the baryon's,\cite{Lu:2017cln,Chen:2017pse}. The diquark content is summarized in Table~\ref{tab:2}:
\begin{center}
    \begin{table}[h!]
\centering
    \caption{Diquark composition of nucleon and $N^*(1535)$, with masses $m_N=1.14$ GeV and $m_{N^*}=1.82$ GeV within the CI.}
\label{tab:2}       
\begin{tabular}{| c | c | c | c | c |}
\hline
& $\;\;\;\;0^+\;\;\;\;$ & $\;\;\;\;1^+\;\;\;\;$ & $\;\;\;\;0^-\;\;\;\;$ & $\;\;\;\;1^-\;\;\;\;$ \\
\hline
$\;N(940)\;$ & $77 \%$ & $22 \%$ & $< 1 \%$ & $<1\%$ \\
$\;N^*(1535)\;$ & $12 \%$ & $<1\%$ & $84 \%$ & $3.5\%$ \\
\hline
\end{tabular} 
\end{table}
\end{center}
As expected, the ground state is mostly constituted by even parity diquarks, while its parity partner has a non-negligible component
of both,~\cite{Lu:2017cln,Chen:2017pse}.
\section{Diquark and $N^*$ transitions}
\begin{figure}
\begin{center}
\vspace*{-6cm}
  \includegraphics[width=120mm]{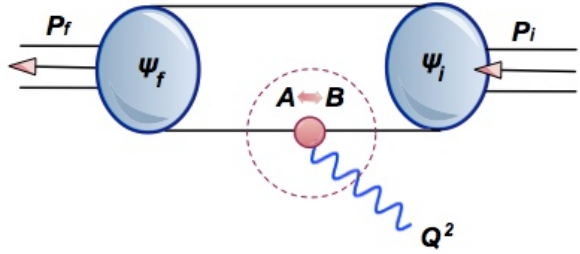}
\vspace*{-6cm}
\caption{Diquark transitions between $N$ and $N^*$.}
\label{figureDT}       
\end{center}
\end{figure}

\begin{figure}
\begin{center}
\vspace*{-5cm}
  \includegraphics[width=120mm]{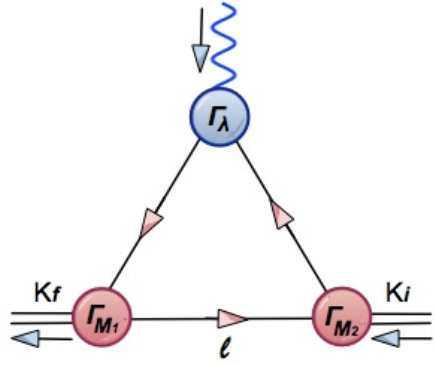}
\vspace*{-5cm}
\caption{Triangle diagram for the diquark transitions.}
\label{fig3}       
\end{center}
\end{figure}

Transition currents for $\gamma^*N\to N^*(1535)$ and elastic processes \cite{Wilson:2011aa,Ramalho:2016buz} are:
\begin{eqnarray}
\nonumber
\mathcal{J}_\mu^{NN^*}&=&ie u_+^{N^*}(P_f)\left[\gamma_\mu^T F_{1^*}(Q^2)+\frac{\sigma_{\mu\nu}Q_\nu}{M_N+M_{N^*}}F_{2^*}(Q^2) \right]\gamma_5 u_+^N(Pi),\\
\mathcal{J}_\mu^{B}&=&ie u_+^B(P_f)\left[\gamma_\mu F_1(Q^2)+\frac{\sigma_{\mu\nu}Q_\nu}{2M_{N^*}}F_2(Q^2) \right]u_+^B(Pi).
\end{eqnarray}
In our picture, several intermediate diquark transitions are required to evaluate $\gamma^*N\to N^*(1535)$ transition. Additionally, one must compute the elastic process $\gamma^*N^*(1535)\to N^*(1535)$ to obtain the canonical normalization constant which ensures charge conservation. Intermediate transitions are shown in figures (\ref{figureDT},\ref{fig3}).

Diquark type $A,B$ could be $0^\pm,1^\pm$, depending on the transition one needs to evaluate. In particular, $2\times4=8$ intermediate transitions are required to compute $\gamma^*N\to N^*(1535)$, and $4\times4=16$ intermediate transitions for the elastic process $\gamma^*N^*(1535)\to N^*(1535)$. Evaluation of the case when the photon hits the quark is also necessary.

\subsection{Quark-photon vertex}
The remaining ingredient is the quark-photon vertex. It is the photon which probes the internal structure of the hadron, therefore
a proper description is required. A lot of work has been invested in constructing such a vertex which obeys all the key requirements
of QED has been, see for example~\cite{Curtis:1990zs,Bashir:1999bd,Kizilersu:2009kg,Bashir:2011vg,Bashir:2011dp,Qin:2013mta}.
For the CI, our task is simpler. We solve the inhomogeneous BSE for the quark-photon vertex:
\begin{equation}
    \Gamma_{\mu}^{RP}(Q)= \gamma_{\mu} - \frac{16 \pi}{3} \, \frac{\alpha_{\rm IR}}{m_G^2}
 \int \frac{d^4q}{(2 \pi)^4} \gamma_{\alpha} \chi_{\mu}(q_+,q)
 \gamma_{\alpha} \,. \label{eq:eqvertex}
\end{equation}
We write $\Gamma_{\mu}^{RP}(Q) = P_L(Q^2) \gamma_\mu^L + P_T(Q^2) \gamma_\mu^T$ ($\gamma_\mu^L+\gamma_\mu^T=\gamma_\mu$). $P_L(Q^2)=1$ and the transverse part, $P_T(Q^2)$, exhibits a pole at $Q^2=-m_\rho^2$.  Additionally, an anomalous electromagnetic moment is then added:
\begin{equation}
    \Gamma_\mu^{AM}(Q) = \frac{\zeta}{2M}\sigma_{\mu\nu}Q_\nu e^{-Q^2/4M^2},\;\zeta = 1/2\;.
\end{equation}
As explained in \cite{Wilson:2011aa}, in the presence of dynamical chiral symmetry breaking, a dressed light-quark possesses a large anomalous electromagnetic moment, which has no effect on the elastic form factor of the scalar diquark but it does change the form factors of the axial-vector diquarks.

\section{Remarks}
We have shown progress towards the computation of $\gamma^*N\to N^*(1535)$ form factor, in a symmetry preserving vector $\times$ vector contact interaction. CI gives us the unique ability to identify those features of hadron observables which are sensitive to the running of the coupling and mass functions in QCD. It provides crucial benchmarks which can be compared and contrasted with QCD connected studies.
\\
\\
The procedure we have described is quite general and can be applied for other transitions. The nucleon transition form factors for large momentum transfer will be measured at the 12 GeV upgrade of the Jefferson Laboratory. Therefore, it is timely to have the predictions ready for the relevant form factors.




\end{document}